\documentclass[aps,prd,twocolumn,superscriptaddress]{revtex4}
\usepackage{epsfig,epsf}
\usepackage{amsmath}
\usepackage{amsthm}
\usepackage{amsfonts}
\usepackage{amssymb}
\usepackage{dsfont}
\usepackage{multirow}
\usepackage{appendix}
\usepackage{slashed}
\usepackage[active]{srcltx}
\usepackage{psfrag}

\begin{document}

\title{The nonet of the light scalar tetraquarks: the mesons $a_0(980)$ and $%
K_{0}^{\ast }(800)$}
\date{\today}
\author{S.~S.~Agaev}
\affiliation{Institute for Physical Problems, Baku State University, Az--1148 Baku,
Azerbaijan}
\author{K.~Azizi}
\affiliation{Department of Physics, Do\v{g}u\c{s} University, Acibadem-Kadik\"{o}y, 34722
Istanbul, Turkey}
\author{H.~Sundu}
\affiliation{Department of Physics, Kocaeli University, 41380 Izmit, Turkey}

\begin{abstract}
The spectroscopic parameters and partial decay widths of the light mesons $%
a_0(980) $ and $K_{0}^{\ast}(800)$ are calculated by treating them as scalar
diquark-antidiquark states. The masses and couplings of the mesons are found
in the framework of QCD two-point sum rule approach. The widths of the decay
channels $a_0(980) \to \eta \pi$ and $a_0(980) \to K \bar{K}$, and $%
K_{0}^{\ast}(800) \to K^{+} \pi^{-}$ and $K_{0}^{\ast}(800) \to K^{0}
\pi^{0} $ are evaluated using QCD sum rules on the light-cone and technical
tools of the soft meson approximation. Our results for the mass of the
mesons $m_{a_0}=991^{+29}_{-27} \ \mathrm{MeV}$ and $m_{K^{%
\ast}}=767^{+38}_{-29} \ \mathrm{MeV}$, as well as their total width $\Gamma
_{\mathrm{a_0}}=62.01\pm 14.37\ \mathrm{MeV}$ and $\Gamma _{\mathrm{%
K_0^{\ast}}}=401.1\pm 87.1\ \mathrm{MeV}$ are compared with last
experimental data.
\end{abstract}

\maketitle


\section{Introduction}

\label{sec:Int} 
The experimental investigation of the light scalar mesons and theoretical
interpretation of obtained data remains one of intriguing problems in high
energy physics. Experimental information on parameters of these particles
suffers from large uncertainties: Their masses and widths are sometimes
known with $\sim 100\ \mathrm{MeV}$ accuracy \cite{Patrignani:2016xqp}. The
status of some of these particles is still unclear, even their existence is
under question.

The theoretical interpretations of light scalars also meet with well-known
troubles. Really, the nonet of scalar particles in the conventional
quark-antiquark model of mesons may be realized as $1{}^{3}P_{0}$ states.
The masses of these scalars, in accordance with various model calculations
are higher than $1\ \mathrm{GeV}$. In fact, the isoscalar mesons $%
f_{0}(1370) $ and $f_{0}(1710)$, the isovector $a_{0}(1450)$ or isospinor $%
K_{0}^{\ast }(1430)$ states were identified as members of the $1{}^{3}P_{0}$
multiplet. But masses of the mesons from the light scalar nonet lie below $1\
\mathrm{GeV} $. Therefore, during a long time the broad scalar resonances $%
f_{0}(500)$ and $K_{0}^{\ast }(800)$, relatively narrow states $f_{0}(980)$
and $a_{0}(980)$ are subject of controversial theoretical hypothesis and
suggestions. The main idea behind attempts to explain unusual features of
these states is an assumption about four-quark (diquark-antidiquark or
meson-meson) nature of these mesons \cite%
{Jaffe:1976ig,Weinstein:1982gc,Weinstein:1990gu}. Within this scheme quantum
numbers and low masses, as well as mass hierarchy inside of the light nonet
seem receive reasonable explanations. The present-day physics of the light
scalars consists of different ideas, models and theories. The comprehensive
information on these issues can be found in the review articles \cite%
{Amsler:2004ps,Bugg:2004xu,Jaffe:2004ph,Klempt:2007cp}.

The diquark-antidiquark picture allows one to answer essential questions
about internal organization of light scalar mesons, and calculate
spectroscopic parameters and decay width of these particles \cite%
{Jaffe:1976ig,Maiani:2004uc,Hooft:2008we}. In this model the scalar mesons
emerge as the nonet of particles composed of four valence quarks. Within the
nonet the $SU_{f}(3)$ flavor octet and singlet states may mix to create the
physical mesons $f_{0}(500)$ and $f_{0}(980)$. The situation here is similar
to the well-known mixing phenomenon in the $\eta -\eta ^{\prime }$ system of
the pseudoscalar mesons. The other two scalar particles $a_{0}(500)$ and $%
K_{0}^{\ast }(800)$ may be identified with the isotriplet and isospinor
members of the light multiplet. A model of the scalar mesons above and below
$1\ \mathrm{GeV}$ was proposed in Ref.\ \cite{Close:2002zu}, in which the
heavy nonet is the conventional $q\overline{q}$ nonet mixed with the
glueball, whereas the light nonet has a four-quark composition with the
diquark-antidiquark or molecule-like structures. An interesting suggestion
about the structure of the scalar mesons was proposed recently in Ref.\ \cite%
{Kim:2017yvd}. In this picture not only light mesons but also the heavy ones
are collected into two nonet of the scalar particles with
diquark-antidiquark structure: The physical mesons are mixtures of the spin-$%
0$ diquarks from ($\overline{\mathbf{3}}_{c},\overline{\mathbf{3}}_{f}$)
representation with spin-$1$ diquarks from ($\mathbf{6}_{c},\ \overline{%
\mathbf{3}}_{f})$ representation of the color-flavor group.

The diquark-antidiquark model allowed one to calculate parameters of the
light scalars and explore their strong and electromagnetic decay channels.
To this end, different calculational schemes and methods were used. Thus,
the masses of the $f_{0}(500)$, $f_{0}(980)$, $a_{0}(980)$ and $K_{0}^{\ast
}(800)$ mesons were calculated in Ref.\ \cite{Ebert:2008id} in the framework
of the relativistic diquark-antidiquark approach and nice agreements with
the data were found. In the context of the four-quark Bethe-Salpeter
equation the same problem was addressed in Ref.\ \cite{Eichmann:2015cra}.
The two-pseudoscalar and two-photon decays of the mesons from the light
scalar nonet were studied in Ref.\ \cite{Giacosa:2006rg}.

Intensive investigations of the light scalar mesons were performed using QCD
sum rules method \cite%
{Latorre:1985uy,Narison:1986vw,Brito:2004tv,Wang:2005cn,
Chen:2007xr,Sugiyama:2007sg,Kojo:2008hk,Zhang:2009qb,Wang:2015uha}. In these
papers apart from the pure diquark-antidiquarks the light scalars were
treated also as mixtures of diquark-antidiquarks with different flavor
structures or as superpositions of diquark-antidiquark and quark-antiquark
components. The aforementioned modification were introduced by the authors
mainly to achieve an agreement between theoretical predictions and
experimental data. The sum rules with inclusion of instanton effects were
employed in Ref.\ \cite{Zhang:2009qb} to evaluate masses of the scalar
mesons above $1~\mathrm{GeV}$. It was demonstrated, that instanton effects
separate the mesons' masses from each other removing the degeneracy of the
conventional sum rules predictions.

In our work \cite{Agaev:2017cfz} we calculated the masses of the mesons $%
f_{0}(500)$ and $f_{0}(980)$ by considering them as states composed of
scalar diquarks. We took into account the mixing of flavor octet and singlet
diquark-antidiquarks that create the physical mesons and, at the same time,
neglected their possible mixing with tetraquarks built of the spin-1
diquarks. Obtained in this work predictions for the masses of the scalar
mesons are in reasonable agreement with existing data. The mixing of the
flavor octet and singlet diquark-antidiquarks used in Ref.\ \cite%
{Agaev:2017cfz} to calculate spectroscopic parameters of the mesons $%
f_{0}(500)$ and $f_{0}(980)$ had important consequences for studying of
their decay channels. Indeed, without octet-singlet mixing the decays of
different scalar mesons proceed through different mechanisms. For example,
the process $f_{0}(980)\rightarrow K\overline{K}$ is the superallowed
Okubo-Zweig-Iizuka (OZI) decay, whereas $f_{0}(980)\rightarrow \pi \pi $ can
proceed due to one gluon exchange \cite{Brito:2004tv}. The octet-singlet
mixing allows one to treat all of the light scalar mesons' decay channels
employing the OZI mechanism, and explain differences in their partial widths
by the mixing parameters. The decays of the $f_{0}(500)$ and $f_{0}(980)$
mesons in this framework were evaluated in Ref. \cite{Agaev:2018}.

The present work is an extension of our previous studies devoted to
spectroscopy and decay properties of the light scalar mesons \cite%
{Agaev:2017cfz,Agaev:2018}. We treat them as diquark-antidiquark states
composed of the scalar diquarks by ignoring their possible mixing with
spin-1 diquarks. We calculate the spectroscopic parameters of the mesons $%
a_{0}(980)$ and $K_{0}^{\ast }(800)$, evaluate their partial decay widths
and, as a result, total widths of these particles. All investigations are
performed using QCD sum rule method: In order to calculate the mass and
coupling of the mesons we employ QCD two-point sum rule approach by
including into analysis quark, gluon and mixing vacuum condensates up to
dimension ten \cite{Shifman1,Shifman2}. The sum rules for the strong
couplings of the vertices $a_{0}(980)\eta \pi ^{0}$, $a_{0}(980)K^{+}K^{-}$,
$K_{0}^{\ast }(800)K^{+}\pi ^{-}$ and $K_{0}^{\ast }(800)K^{0}\pi ^{0}$ are
derived using light-cone sum rule (LCSR) method \cite{Balitsky:1989ry} and
technical tools of the soft-meson approximation \cite{Belyaev:1994zk}, which
was adapted in Ref.\ \cite{Agaev:2016dev} to study tetraquark-meson-meson
vertices. This approach was successfully applied to evaluate strong
couplings and widths of numerous decays involving tetraquarks \cite%
{Agaev:2017lmc,Agaev:2017oay}, including the light axial-vector meson $%
a_{1}(1420)$ \cite{Sundu:2017xct}.

This paper is organized in the following way: In the section \ref{sec:Masses}
we calculate the mass and coupling of the mesons $a_{0}(980)$ and $%
K_{0}^{\ast }(800)$. In the section \ref{sec:Decay1} we derive the sum rules
to evaluate the strong couplings $g_{a\eta \pi }$, $g_{aKK}$, $g_{K^{\ast
}K\pi }$ and $g_{K^{\ast }K^{0}\pi ^{0}}$. The obtained results are utilized
in Sec. \ref{sec:NRes} for numerical evaluation of the strong couplings and
widths of the decays $a_{0}(980)\rightarrow \eta \pi ^{0}$, $%
a_{0}(980)\rightarrow K^{+}K^{-}$, $K_{0}^{\ast }(800)\rightarrow K^{+}\pi
^{-}$ and $K_{0}^{\ast }(800)\rightarrow K^{0}\pi ^{0}$, and total widths of
the mesons $a_{0}(980)$ and $K_{0}^{\ast }(800)$. In Section \ref{sec:CRel}
we discuss obtained results and present our concluding notes.


\section{Mass and coupling of the mesons $a_0(980)$ and $K_{0}^{\ast}(800)$}

\label{sec:Masses}


The mass and coupling of the mesons $a_0(980)$ and $K_{0}^{\ast}(800)$ can
be calculated within QCD two-point sum rule method. We consider here in
details all necessary steps to find the mass and coupling of the $a_0(980)$
meson and provide only final expressions and results for the $%
K_{0}^{\ast}(800)$ meson.

The mass and coupling of $a_{0}(980)$ can be extracted from the sum rule
analysis of the two-point correlation function%
\begin{equation}
\Pi (p)=i\int d^{4}xe^{ipx}\langle 0|\mathcal{T}\{J(x)J^{^{\dagger
}}(0)\}|0\rangle ,  \label{eq:CF1}
\end{equation}%
where $J(x)$ is the interpolating current to the $a_{0}(980)$ meson. In the
diquark-antidiquark model it can be written in the following form
\begin{eqnarray}
&&J(x)=\frac{\epsilon \widetilde{\epsilon }}{\sqrt{2}}\left[ \left(
u_{a}^{T}C\gamma _{5}s_{b}\right) \left( \overline{u}_{d}\gamma _{5}C%
\overline{s}_{e}^{T}\right) \right.  \notag \\
&&\left. -\left( d_{a}^{T}C\gamma _{5}s_{b}\right) \left( \overline{d}%
_{d}\gamma _{5}C\overline{s}_{e}^{T}\right) \right] ,  \label{eq:Current1}
\end{eqnarray}%
where $C$ is the charge conjugation operator. Here we also use the
short-hand notation $\epsilon \widetilde{\epsilon }=\epsilon _{abc}\epsilon
_{dec}$ with $a,b,c,d$ and $e$ being the color indices. Let us note that we
use the conventional two-point sum rules neglecting the possible instanton
effects in the correlation function $\Pi (p)$.

In accordance with standard prescriptions of the sum rule computations the
correlation function $\Pi (p)$ should be found by employing both the
physical parameters of the $a_{0}(980)$ meson, i. e. its mass $m_{a_{0}}$
and coupling $f_{a_{0}}$ and in terms of the light-quark propagators, and as
a result, in terms of various quark, gluon and mixed vacuum condensates. By
matching the obtained results and benefiting from the assumption on the
quark-hadron duality it is possible to extract sum rules and evaluate the
physical parameters of interest.

In the case under consideration the physical side of the sum rule takes the
simple form%
\begin{equation}
\Pi ^{\mathrm{Phys}}(p)=\frac{\langle 0|J|a_{0}(p)\rangle \langle
a_{0}(p)|J^{\dagger }|0\rangle }{m_{a_{0}}^{2}-p^{2}}+\ldots ,
\label{eq:PhysSide1}
\end{equation}%
because the $a_{0}(980)$ meson is the ground-state particle: The
contributions coming from the excited and continuum states are shown in Eq.\
(\ref{eq:PhysSide1}) by dots. To express $\Pi ^{\mathrm{Phys}}(p)$ in terms
of the parameters $m_{a_{0}}$ and $f_{a_{0}}$ we introduce the matrix element%
\begin{equation}
\langle 0|J|a_{0}(p)\rangle =f_{a_{0}}m_{a_{0}},  \label{eq:ME1}
\end{equation}%
and get%
\begin{equation*}
\Pi ^{\mathrm{Phys}}(p)=\frac{f_{a_{0}}^{2}m_{a_{0}}^{2}}{m_{a_{0}}^{2}-p^{2}%
}+\ldots .
\end{equation*}%
Effect of the excited states and continuum on the $\Pi ^{\mathrm{Phys}}(p)$
can be suppressed by means of the Borel transformation which yields%
\begin{equation}
\mathcal{B}\Pi ^{\mathrm{Phys}%
}(p)=f_{a_{0}}^{2}m_{a_{0}}^{2}e^{-m_{a_{0}}^{2}/M^{2}}+\ldots ,
\label{eq:BorelTr1}
\end{equation}%
where $M^{2}$ is the Borel parameter. In Eq.\ (\ref{eq:BorelTr1}) by dots we
again denote contributions of the excited states and continuum which will be
subtracted from Borel transformation of $\Pi ^{\mathrm{OPE}}(p)$ to derive
the required sum rules.

The $\Pi ^{\mathrm{OPE}}(p)$ that constitutes the second part of the sum
rule's equality is obtained from Eq.\ (\ref{eq:CF1}) using the explicit
expression for the interpolating current $J(x)$ and contracting the relevant
quarks fields. As a result for $\Pi ^{\mathrm{OPE}}(p)$ we find%
\begin{eqnarray}
&&\Pi ^{\mathrm{OPE}}(p)=i\int d^{4}xe^{ipx}\frac{\epsilon \widetilde{%
\epsilon }\epsilon ^{\prime }\widetilde{\epsilon }^{\prime }}{2}\left\{
\mathrm{Tr}\left[ \gamma _{5}\widetilde{S}_{s}^{e^{\prime }e}(-x)\gamma
_{5}\right. \right.  \notag  \label{eq:OPE1} \\
&&\left. \left. \times S_{u}^{d^{\prime }d}(-x)\right] \mathrm{Tr}\left[
\gamma _{5}\widetilde{S}_{u}^{aa^{\prime }}(x)\gamma _{5}S_{s}^{bb^{\prime
}}(x)\right] +(u\leftrightarrow d)\right\} .  \notag \\
&&
\end{eqnarray}%
In the expression above
\begin{equation*}
\widetilde{S}_{s(q)}(x)=CS_{s(q)}^{T}(x)C,
\end{equation*}%
where $S_{s(q)}(x)$ are the $s$ and $q=u,d$ quarks' propagators
\begin{eqnarray}
&&S_{q}^{ab}(x)=i\delta _{ab}\frac{\slashed x}{2\pi ^{2}x^{4}}-\delta _{ab}%
\frac{m_{q}}{4\pi ^{2}x^{2}}-\delta _{ab}\frac{\langle \overline{q}q\rangle
}{12}  \notag \\
&&+i\delta _{ab}\frac{\slashed xm_{q}\langle \overline{q}q\rangle }{48}%
-\delta _{ab}\frac{x^{2}}{192}\langle \overline{q}g_{s}\sigma Gq\rangle
\notag \\
&&-i\frac{g_{s}G_{ab}^{\alpha \beta }}{32\pi ^{2}x^{2}}\left[ \slashed x{%
\sigma _{\alpha \beta }+\sigma _{\alpha \beta }}\slashed x\right] +i\delta
_{ab}\frac{x^{2}\slashed xm_{q}}{1152}\langle \overline{q}g_{s}\sigma
Gq\rangle  \notag \\
&&-i\delta _{ab}\frac{x^{2}\slashed xg_{s}^{2}\langle \overline{q}q\rangle
^{2}}{7776}-\delta _{ab}\frac{x^{4}\langle \overline{q}q\rangle \langle
g_{s}^{2}G^{2}\rangle }{27648}+...  \label{eq:Qprop}
\end{eqnarray}%
In the present work we calculate the correlation function by taking into
account nonperturbative terms up to dimension ten.

The Borel transform of the correlator $\mathcal{B}\Pi ^{\mathrm{OPE}}(p)=$ $%
\Pi ^{\mathrm{OPE}}(M^{2})$ can be calculated using either the spectral
density $\rho (s)$ which is proportional to imaginary part of $\Pi ^{\mathrm{%
OPE}}(p)$ or by applying the Borel transformation directly to $\Pi ^{\mathrm{%
OPE}}(p)$. If necessary, $\Pi ^{\mathrm{OPE}}(M^{2})$ may be computed
utilizing both of these approaches. These routine operations were explained
numerously in existing literature, therefore we do not concentrate on these
questions here. The obtained expression for $\Pi ^{\mathrm{OPE}}(M^{2})$ has
to be equated to Eq.\ (\ref{eq:BorelTr1}), and one also has to perform the
continuum subtraction. After these manipulations we find the following sum
rule%
\begin{equation}
f_{a_{0}}^{2}m_{a_{0}}^{2}e^{-m_{a_{0}}^{2}/M^{2}}=\Pi ^{\mathrm{OPE}%
}(M^{2},s_{0}),  \label{eq:SR1}
\end{equation}%
where $\Pi ^{\mathrm{OPE}}(M^{2},s_{0})$ is now the continuum subtracted
correlation function. In Eq.\ (\ref{eq:SR1}) $s_{0}$ is the continuum
threshold parameter: It separates from each other contribution of the
ground-state term and effects due to excited states and continuum. The
second sum rule is derived by applying operator $d/d(-1/M^{2})$ to Eq.\ (\ref%
{eq:SR1})%
\begin{equation}
f_{a_{0}}^{2}m_{a_{0}}^{4}e^{-m_{a_{0}}^{2}/M^{2}}=\widetilde{\Pi }^{\mathrm{%
OPE}}(M^{2},s_{0}),  \label{eq:SR2}
\end{equation}%
where $\widetilde{\Pi }^{\mathrm{OPE}}(M^{2},s_{0})=d/d(-1/M^{2})\Pi ^{%
\mathrm{OPE}}(M^{2},s_{0})$. These two sum rules can be employed to evaluate
the parameters $m_{a_{0}}$ and $f_{a_{0}}$:%
\begin{equation}
m_{a_{0}}^{2}=\frac{\widetilde{\Pi }^{\mathrm{OPE}}(M^{2},s_{0})}{\Pi ^{%
\mathrm{OPE}}(M^{2},s_{0})},  \label{eq:Mass}
\end{equation}%
and
\begin{equation}
f_{a_{0}}^{2}=\frac{e^{m_{a_{0}}^{2}/M^{2}}}{m_{a_{0}}^{2}}\Pi ^{\mathrm{OPE}%
}(M^{2},s_{0}).  \label{eq;Coupling}
\end{equation}

The sum rules for the parameters of the meson $K_{0}^{\ast }(800)$ can be
found by the same manner. Differences in this case are connected with the
interpolating current of $K_{0}^{\ast }(800)$ defined by the expression
\begin{equation}
J^{K^{\ast }}(x)=\epsilon \widetilde{\epsilon }\left( u_{a}^{T}C\gamma
_{5}d_{b}\right) \left( \overline{u}_{d}\gamma _{5}C\overline{s_{e}}%
^{T}\right) ,  \label{eq:Current2}
\end{equation}%
and with the matrix element%
\begin{equation}
\langle 0|J^{K^{\ast }}|K_{0}^{\ast }(p)\rangle =f_{K^{\ast }}m_{K^{\ast }},
\label{eq:ME2}
\end{equation}%
where $m_{K^{\ast }}$ and $f_{K^{\ast }}$ are the mass and coupling of the
state $K_{0}^{\ast }(800)$. The phenomenological side of the sum rule after
evident replacements is given by Eq.\ (\ref{eq:BorelTr1}), whereas the $\Pi
_{K^{\ast }}^{\mathrm{OPE}}(p)$ takes the following form%
\begin{eqnarray}
&&\Pi _{K^{\ast }}^{\mathrm{OPE}}(p)=i\int d^{4}xe^{ipx}\epsilon \widetilde{%
\epsilon }\epsilon ^{\prime }\widetilde{\epsilon }^{\prime }\mathrm{Tr}\left[
\gamma _{5}\widetilde{S}_{s}^{e^{\prime }e}(-x)\right.  \notag \\
&&\left. \times \gamma _{5}S_{u}^{d^{\prime }d}(-x)\right] \mathrm{Tr}\left[
\gamma _{5}\widetilde{S}_{u}^{aa^{\prime }}(x)\gamma _{5}S_{d}^{bb^{\prime
}}(x)\right] .  \label{eq:OPE2}
\end{eqnarray}%
The remaining operations are standard and do not differ from ones described
above in the case of the $a_{0}(980)$ meson.

The numerical computations require to specify values of various parameters
that enter to the quark propagators,  and, as a result, to the sum rules for the mass
and coupling. Among them the vacuum expectation values of the quark, gluon
and mixed local operators are important ones:
\begin{eqnarray}
&&\langle \bar{q}q\rangle =-(0.24\pm 0.01)^{3}\ \mathrm{GeV}^{3},\ \langle
\bar{s}s\rangle =0.8\ \langle \bar{q}q\rangle ,  \notag \\
&&m_{0}^{2}=(0.8\pm 0.1)\ \mathrm{GeV}^{2},\ \langle \overline{q}g_{s}\sigma
Gq\rangle =m_{0}^{2}\langle \overline{q}q\rangle ,  \notag \\
&&\langle \overline{s}g_{s}\sigma Gs\rangle =m_{0}^{2}\langle \bar{s}%
s\rangle ,  \notag \\
&&\langle \frac{\alpha _{s}G^{2}}{\pi }\rangle =(0.012\pm 0.004)\,\mathrm{GeV%
}^{4}.  \label{eq:Param}
\end{eqnarray}%
These condensates enter to the propagator of a light quark and have
different dimensions. The terms $\langle \overline{q}g_{s}\sigma Gq\rangle $%
, $\langle \overline{s}g_{s}\sigma Gs\rangle $ shown in Eq.\ (\ref{eq:Param}%
) as well as other ones $\sim \langle \overline{q}q\rangle ^{2}$, $\sim
\langle \overline{q}q\rangle \langle g_{s}^{2}G^{2}\rangle $ are obtained
using the factorization hypothesis of the higher dimension condensates.
However, the factorization assumption is not precise and its violation is
stronger for higher dimension condensates (see Ref.\ \cite{Ioffe:2005ym}).
For dimension ten condensates even the order of magnitude of such a
violation is unclear. But here we employ this assumption by ignoring
possible theoretical uncertainties generated by its violation.

In the present work we neglect the masses of the $u$ and $d$ quarks, but set
$m_{s}\neq 0$ and use in calculations $m_{s}=128\pm 10~\mathrm{MeV}$. Our
expressions depend also on auxiliary parameters $M^{2}$ and $s_{0}$ the
choice of which has to satisfy standard restrictions. Thus, we determine the
upper limit $M_{\mathrm{max}}^{2}$ of the working window $M^{2}\in \lbrack
M_{\mathrm{min}}^{2},\ M_{\mathrm{max}}^{2}]$ by requiring fulfillment of
the condition imposed on the pole contribution
\begin{equation}
\mathrm{PC}=\frac{\Pi (M_{\mathrm{max}}^{2},\ s_{0})}{\Pi (M_{\mathrm{max}%
}^{2},\ \infty )}>0.10.  \label{eq:Rest1}
\end{equation}

The lower bound of the Borel parameter $M_{\mathrm{min}}^{2}$ is fixed from
convergence of the operator product expansion (OPE). By quantifying this
constraint we require that a contribution of the last term in OPE should be
around of $5\%$, i. e.
\begin{equation}
\frac{\Pi ^{\mathrm{Dim10}}(M_{\mathrm{min}}^{2},\ s_{0})}{\Pi (M_{\mathrm{%
min}}^{2},\ s_{0})}\approx 0.05,  \label{eq:Rest2}
\end{equation}%
has to be obeyed. Another restriction to $M_{\mathrm{min}}^{2}$ is connected
with the perturbative contribution to sum rules. In the present work we
apply the following criterion: at the lower bound of $M^{2}$ the
perturbative contribution has to constitute more than $70\%$ part of the
full result.

Boundaries of $s_{0}$ are fixed by analyzing the pole contribution to get
its greatest accessible values. Minimal dependence of extracted quantities
on $M^{2}$ while varying $s_{0}$ is another constraint that has to be
imposed when choosing a region for this parameter. Performed analyses lead
to the following working windows for $M^{2}$ and $s_{0}$:
\begin{equation}
M^{2}\in \lbrack 1.1,\ 1.4]\ \mathrm{GeV}^{2},\ s_{0}\in \lbrack 1.7,\ 1.9]\
\mathrm{GeV}^{2}.  \label{eq:Wind1}
\end{equation}%
In these regions all of constraints imposed on the correlation function are
satisfied. In fact, at $M_{\mathrm{max}}^{2}$ the pole contribution \textrm{%
PC } equals to $0.115,$ whereas at $M_{\mathrm{min}}^{2}$ it amounts to $%
78\% $ of the result. In other words, Eq.\ (\ref{eq:Rest1}) determines only
the lower limit for the $\mathrm{PC}$: in the full interval for $M^{2}$ the
pole contribution is large which should lead to reliable sum rules'
predictions. At the minimal allowed value of the Borel parameter
contribution of $\mathrm{Dim10}$ term constitutes up to $5.5\%$ of the whole
result. And perturbative component of the correlation function $\Pi (M_{%
\mathrm{min}}^{2},\ s_{0})$ forms its no less than $0.71$ part.

In Figs.\ \ref{fig:Mass} and \ref{fig:Coupl} we depict the sum rules results
for the mass and coupling of the $a_{0}(980)$ state as functions of the
Borel and continuum threshold parameters. It is seen, that predictions for
the mass and coupling are rather stable against varying of both $M^{2}$ and $%
s_{0}$. In the case of the mass the stability of the result has standard
explanation: In fact, the sum rule for the mass $m_{a_{0}}$ depends on the
ratio of the correlation function and its derivative (\ref{eq:Mass}), where
uncertainties to a great extend cancel rendering the mass very stable in the
working regions of $M^{2}$ and $s_{0}$. The stability of the coupling may be
attributed to the fact that interpolating current $J(x)$ contains only light
diquarks (antidiquarks) $\epsilon _{abc}q_{a}^{T}C\gamma _{5}q_{b}^{\prime }$
in color triplet, flavor antisymmetric and spin $0$ state, and which leads
to stable predictions.

\begin{widetext}

\begin{figure}[h!]
\begin{center} \includegraphics[%
totalheight=6cm,width=8cm]{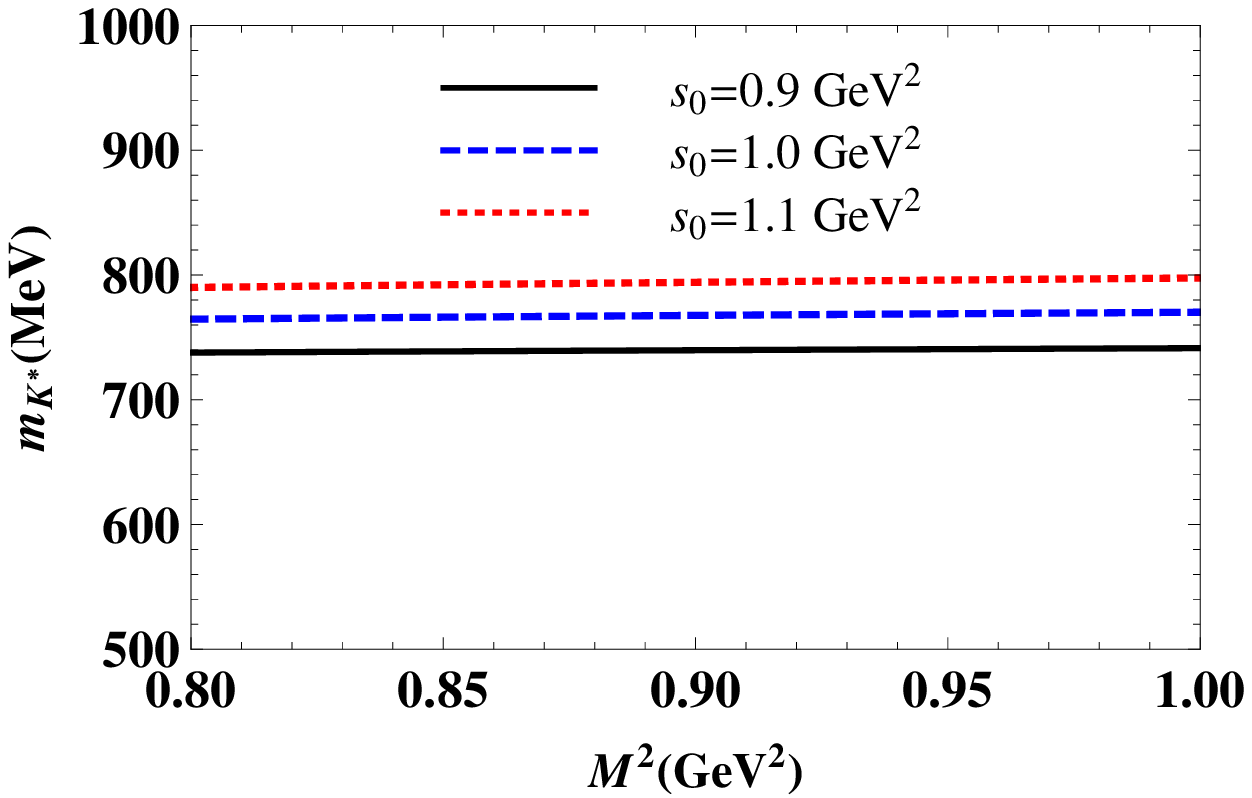}\,\,
\includegraphics[%
totalheight=6cm,width=8cm]{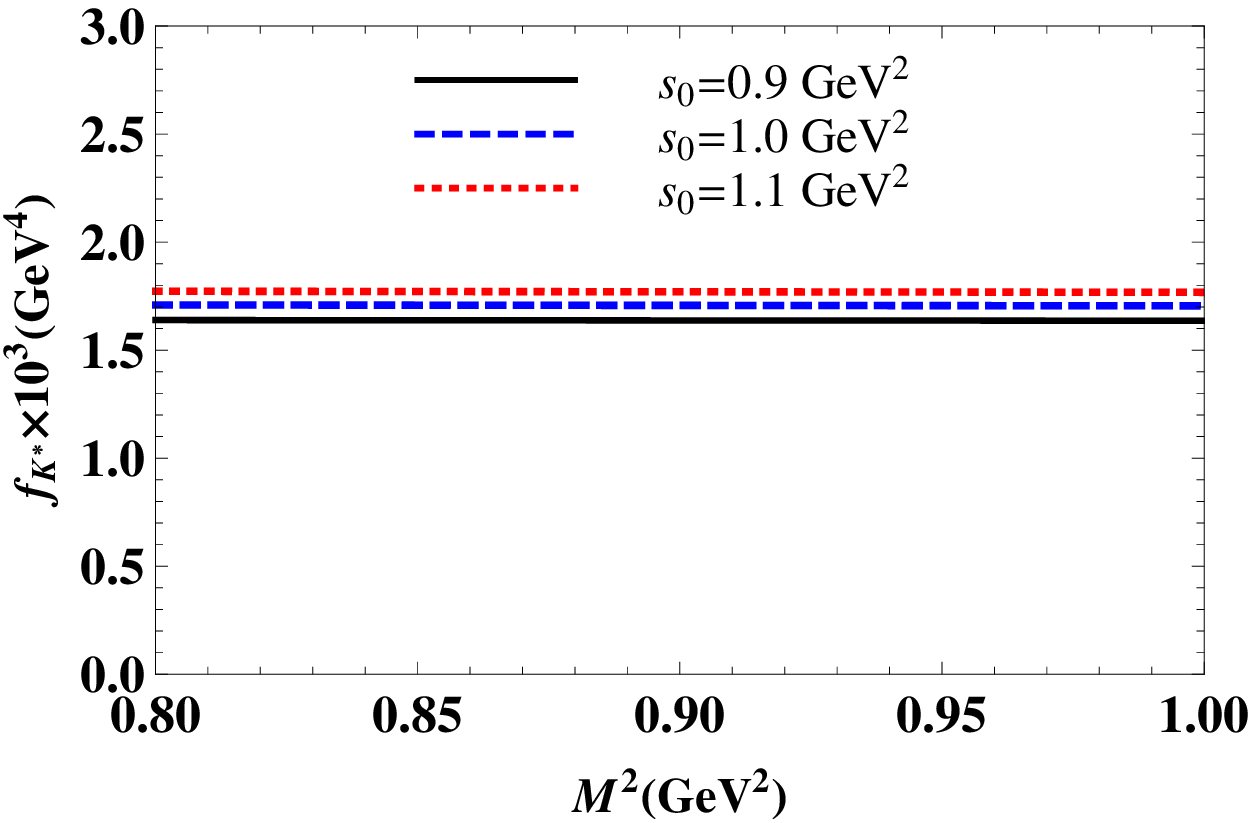}
\end{center}
\caption{ The $m_{K^{\ast }}$ (left panel) and $f_{K^{\ast }}$ (right panel) vs $M^2$ at fixed values of the
continuum threshold parameter $s_0$.}
\label{fig:MassCoupl}
\end{figure}

\end{widetext}

\begin{widetext}

\begin{figure}[h!]
\begin{center} \includegraphics[%
totalheight=6cm,width=8cm]{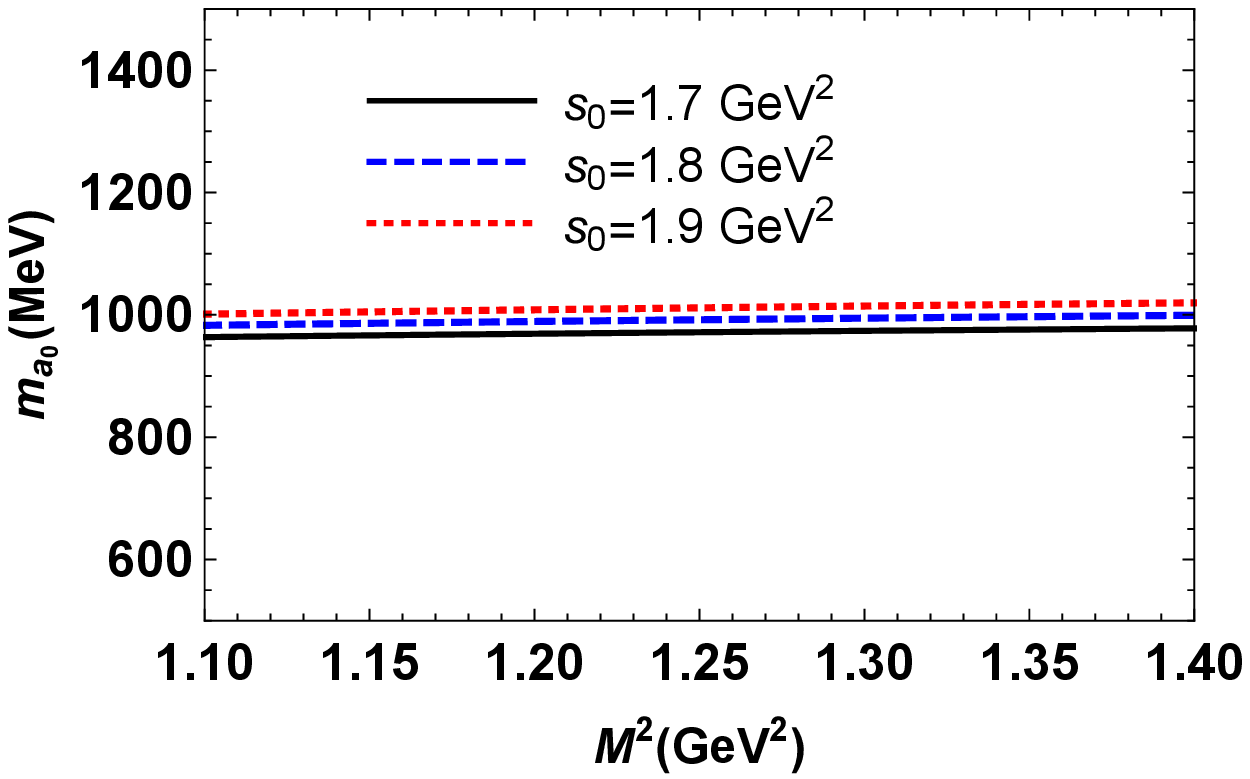}\,\,
\includegraphics[%
totalheight=6cm,width=8cm]{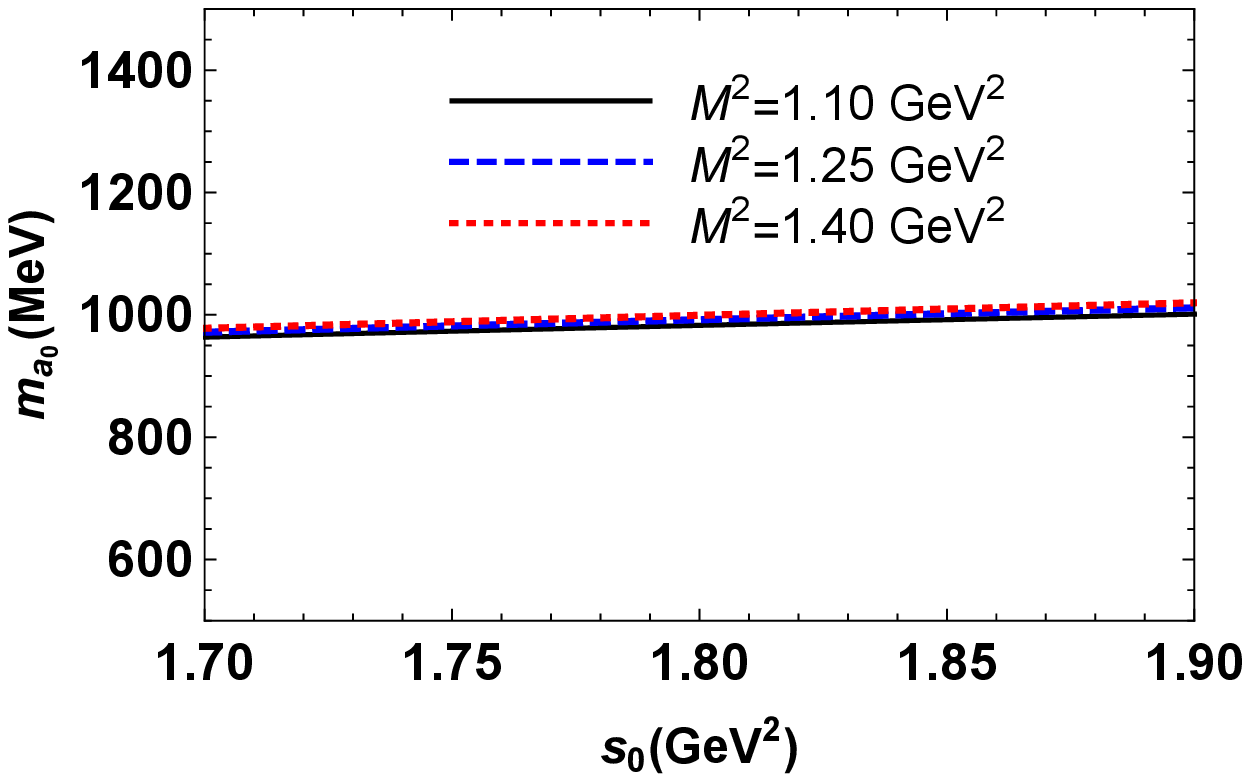}
\end{center}
\caption{ The mass
of the meson $a_0(980)$ as a function of the Borel parameter $M^2$ at fixed $s_0$
(left panel), and as a function of the continuum threshold $s_0$ at fixed $%
M^2$ (right panel).}
\label{fig:Mass}
\end{figure}
\begin{figure}[h!]
\begin{%
center}
\includegraphics[totalheight=6cm,width=8cm]{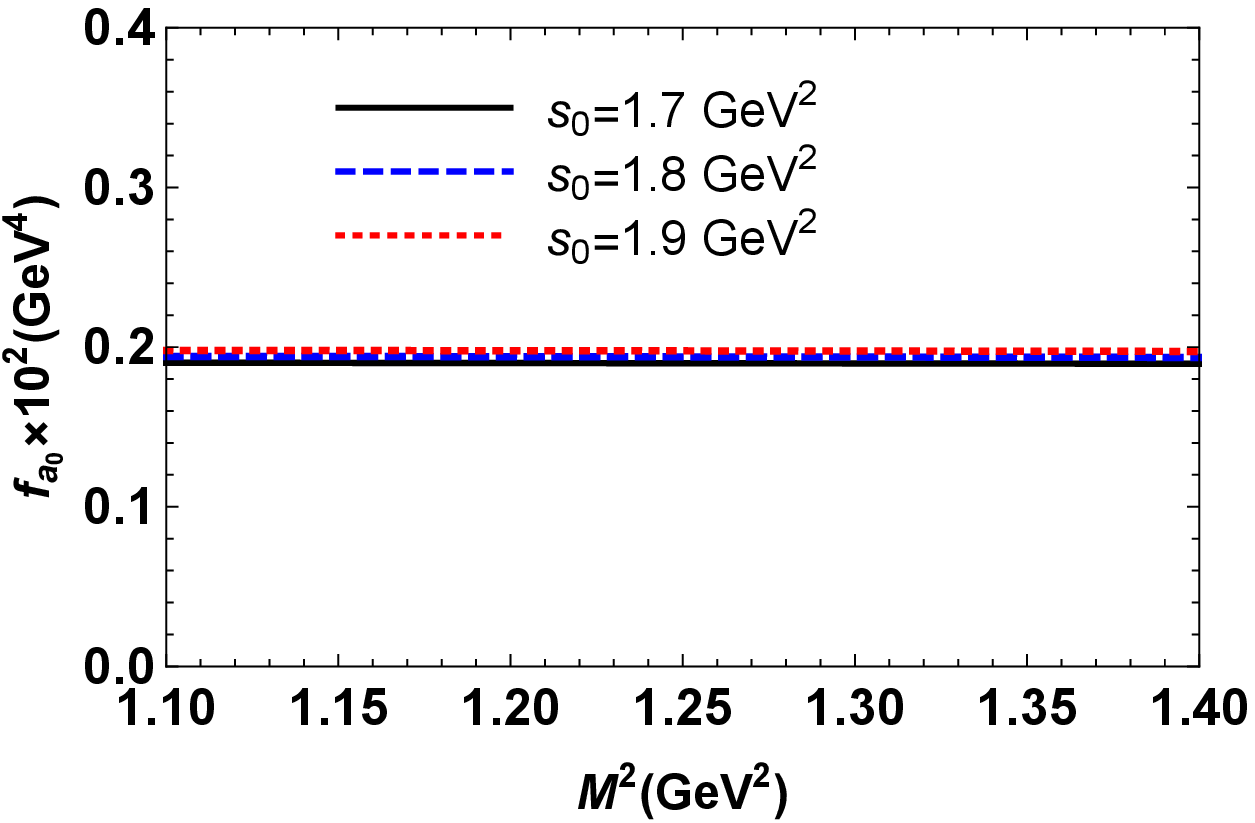}\,\,
\includegraphics[totalheight=6cm,width=8cm]{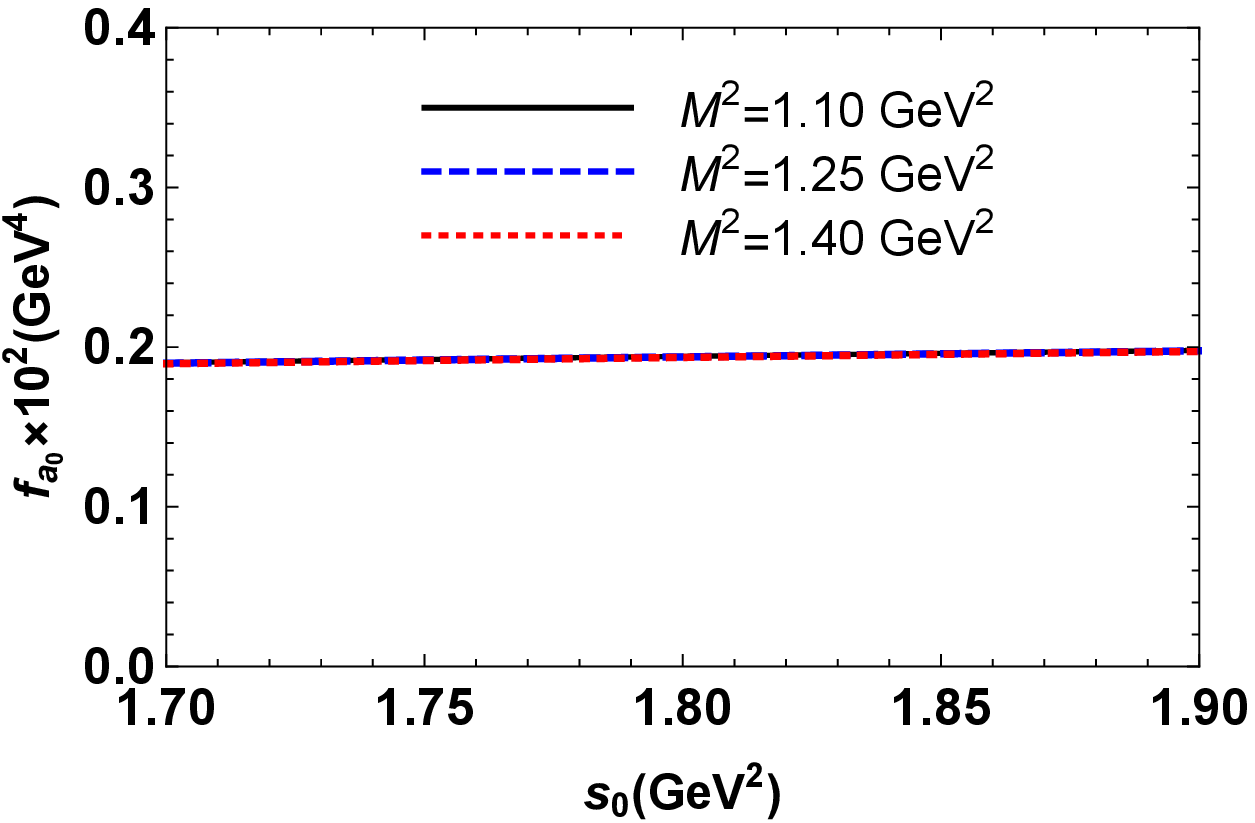} \end{center}
\caption{ The coupling $f_{a_0}$ of the $a_0$ state as a function of $%
M^2$ at fixed $s_0$ (left panel), and of $s_0$ at fixed $M^2$ (right panel).}
\label{fig:Coupl}
\end{figure}

\end{widetext}For $m_{a_{0}}$ and $f_{a_{0}}$ we find:%
\begin{equation}
m_{a_{0}}=991_{-27}^{+29}\ \ \mathrm{MeV},\ f_{a_{0}}=(1.94\pm 0.04)\cdot
10^{-3}\ \mathrm{GeV}^{4}.  \label{eq:MCoupl1}
\end{equation}%
The similar analysis of the sum rules for the mass and coupling of the $%
K_{0}^{\ast }(800)$ meson allows us to find the regions for the Borel and
continuum threshold parameters
\begin{equation}
M^{2}\in \lbrack 0.8,\ 1.0]\ \mathrm{GeV}^{2},\ s_{0}\in \lbrack 0.9,\ 1.1]\
\mathrm{GeV}^{2},  \label{eq:Wind2}
\end{equation}%
which lead to the following predictions:%
\begin{equation}
m_{K^{\ast }}=767_{-29}^{+38}\ \ \mathrm{MeV},\ \ f_{K^{\ast }}=(1.71\pm
0.07)\cdot 10^{-3}\ \mathrm{GeV}^{4}.  \label{eq:MCoupl2}
\end{equation}%
The sum rules predictions for $m_{K^{\ast }}$ and $f_{K^{\ast }}$ are
plotted in Fig.\ \ref{fig:MassCoupl} as functions of the Borel parameter $%
M^{2}$. Their stability on $M^{2}$ including a region $s_{0}<1\ \mathrm{GeV}%
^{2}$ demonstrates correctness of the performed calculations. Our result for
the mass of the $a_{0}(980)$ meson is in a nice agreement with the available
experimental data $m_{a_{0}}=980\pm 20\ \mathrm{MeV}$ \cite%
{Patrignani:2016xqp}. The latest measurement of $m_{K^{\ast }}$ performed by
the BES Collaboration \cite{Ablikim:2010ab} and extracted from the decay $%
J/\psi \rightarrow K_{\mathrm{S}}^{0}K_{\mathrm{S}}^{0}$ $\pi ^{+}\pi ^{-}$
is equal to%
\begin{equation}
m_{K^{\ast }}=826\pm 49_{\ -34}^{\ +49}\ \ \mathrm{MeV.}
\end{equation}%
From the process $J/\psi \rightarrow K\pm K_{\mathrm{S}}^{0}$ $\pi ^{\mp
}\pi ^{0}$ the same collaboration obtained (see, Ref.\ \cite{Ablikim:2010kd}%
)
\begin{equation}
m_{K^{\ast }}=849\pm 77_{\ -14}^{\ +18}\ \ \mathrm{MeV.}
\end{equation}%
As is seen, the experimental data are not precise, and the central values
for $m_{K^{\ast }}$ are higher than our prediction. Nevertheless, within the
experimental and theoretical errors they are compatible with each other. The
mass and coupling of the $a_{0}(980)$ and $K_{0}^{\ast }(800)$ mesons
calculated in the present section will be used as input parameters below to
find their partial decay widths.

\section{Strong decay channels of the $a_{0}(980)$ and $K_{0}^{\ast }(800)$
mesons}

\label{sec:Decay1}

In the light of the obtained results we can determine the kinematically
allowed strong decay channels of the $a_{0}(980)$ and $K_{0}^{\ast }(800)$
mesons. In the present paper we restrict ourselves by studying only $S-$wave
decays of these mesons. It turns out that the dominant $S-$wave strong \
decays of $a_{0}(980)$ are processes $a_{0}(980)\rightarrow \eta \pi ^{0}$
and $a_{0}(980)\rightarrow K^{+}K^{-}$. For the meson $K_{0}^{\ast }(800)$
the decays $K_{0}^{\ast }(800)\rightarrow K^{+}\pi ^{-}$ and $K_{0}^{\ast
}(800)\rightarrow K^{0}\pi ^{0}$ are dominant ones.

These decays proceed through rearrangement of the quarks and antiquarks from
the tetraquark to form two conventional mesons. Mechanisms of these
transformations are not quite clear, but there are interesting models to
explain these phenomena introducing, for instance, a repulsive barrier
between the diquark-antidiquark pair \cite{Esposito:2018cwh}.   
The light cone sum rule method
operates with fundamental quark-gluon degrees of freedom and  uses first
principles of the QCD. In this approach one invokes  only an assumption on
the quark-hadron duality to match the phenomenological and theoretical
expressions of the same correlation function to derive the sum rules for
quantities of interest.

It is instructive to consider the mode $a_{0}(980)\rightarrow \eta \pi ^{0}$
in a detailed manner. In order to calculate the strong coupling $g_{a\eta
\pi }$ we use QCD LCSR method and start from analysis of the correlation
function
\begin{equation}
\Pi (p,q)=i\int d^{4}xe^{ip\cdot x}\langle \pi ^{0}(q)|\mathcal{T}\{J^{\eta
}(x)J^{^{\dagger }}(0)\}|0\rangle ,  \label{eq:CF2}
\end{equation}%
where $J(x)$ and $J^{\eta }(x)$ are the interpolating currents for the $%
a_{0}(980)$ and $\eta $ mesons, respectively. The interpolating current for
the $a_{0}(980)$ is given by Eq.\ (\ref{eq:Current1}).

The situation with the choice of $J^{\eta }(x)$ is more subtle and deserves
some explanations. The system of pseudoscalar mesons $\eta -\eta ^{\prime }$
has a complicated structure. In the world of the exact flavor $SU_{f}(3)$
symmetry the mesons $\eta $ and $\eta ^{\prime }$ can be interpreted as the
octet $\eta _{8}$ and singlet $\eta _{1}$ states of the flavor group,
respectively. But in the real world, where this symmetry is broken the
physical particles are mixtures of the $\eta _{8}$ and $\eta _{1}$ states.
Of course, the mesons $\eta $ and $\eta ^{\prime }$ are predominantly the $%
\eta _{8}$ and $\eta _{1}$ states, nevertheless the mixing phenomenon can
not be ignored. This mixing can be described using the octet-singlet basis.
Alternatively, the same phenomenon can be treated employing the quark-flavor
basis (see, Ref.\ \cite{Feldmann:1999uf} for details)
\begin{equation}
\eta _{q}=\frac{1}{\sqrt{2}}\left( \overline{u}u+\overline{d}d\right) ,\ \ \
\ \eta _{s}=\overline{s}s.  \label{eq:QF}
\end{equation}%
The quark-flavor basis is more convenient to describe the mixing in the $%
\eta -\eta ^{\prime }$ system and investigate different exclusive processes
involving these mesons \cite{Agaev:2014wna}. The reason is that in this
scheme with rather high accuracy the state and coupling mixing are governed
by the same angle, whereas in the $\eta _{1}-\eta _{8}$ basis one has to
introduce two mixing angles for the decay constants.

In the quark-flavor basis the interpolating current of the $\eta $ meson can
be obtained through mixing from the basic currents
\begin{eqnarray}
J_{q}(x) &=&\frac{1}{\sqrt{2}}\left[ \overline{u}(x)i\gamma _{5}u(x)+%
\overline{d}(x)i\gamma _{5}d(x)\right] ,  \notag \\
J_{s}(x) &=&\overline{s}(x)i\gamma _{5}s(x),  \label{eq:BasicCurr}
\end{eqnarray}%
and reads
\begin{equation*}
J^{\eta }(x)=J_{q}(x)\cos \varphi -J_{s}(x)\sin \varphi ,
\end{equation*}%
where $\varphi $ is the mixing angle.

The phenomenological side of the sum rule is obtained by expressing $\Pi
(p,q)$ in terms of the strong coupling $g_{a\eta \pi }$ and physical
parameters of the $a_{0}(980)$ and $\eta $ mesons
\begin{eqnarray}
&&\Pi ^{\mathrm{Phys}}(p,q)=\frac{\langle 0|J^{\eta }|\eta (p)\rangle }{%
p^{2}-m_{\eta }^{2}}\langle \eta (p)\pi ^{0}(q)|a_{0}(p^{\prime })\rangle
\notag \\
&&\times \frac{\langle a_{0}(p^{\prime })|J^{\dagger }|0\rangle }{p^{\prime
2}-m_{a_{0}}^{2}}+\ldots ,  \label{eq:PhysSide2}
\end{eqnarray}%
where $m_{\eta }$ is the mass of $\eta $ the dots being stood for
contributions of excited states. The matrix element $\langle a_{0}(p^{\prime
})|J^{\dagger }|0\rangle $ has been introduced in the previous section, and
the vertex $\langle \eta (p)\pi ^{0}(q)|a_{0}(p^{\prime })\rangle $ can be
written down in the following form
\begin{equation}
\langle \eta (p)\pi ^{0}(q)|a_{0}(p^{\prime })\rangle =g_{a\eta \pi }p\cdot
p^{\prime },  \label{eq:Vertex1}
\end{equation}%
where $g_{a\eta \pi }$ is the coupling corresponding to the strong vertex $%
a_{0}(980)\eta \pi ^{0}$. The last element in Eq.\ (\ref{eq:PhysSide2}) $%
\langle 0|J^{\eta }|\eta (p)\rangle $ is defined by the expression
\begin{equation}
\langle 0|J^{\eta }|\eta (p)\rangle =-\frac{1}{2m_{s}}\left( h_{\eta
}^{q}\cos \varphi -h_{\eta }^{s}\sin \varphi \right)  \label{eq:ME3}
\end{equation}%
and differs from the similar matrix elements of conventional pseudoscalar
mesons: here relevant comments are in order. It is known that the
axial-anomaly modifies the matrix elements of the $\eta $ and $\eta ^{\prime
}$ mesons. Indeed, for $h_{\eta }^{s(q)}$ we have
\begin{equation}
h_{\eta }^{s(q)}=m_{\eta }^{2}f_{\eta }^{s(q)}-\langle 0|\frac{\alpha _{%
\mathrm{s}}}{\pi }G_{\mu \nu }^{A}\widetilde{G}^{A,\mu \nu }|\eta (p)\rangle
,
\end{equation}%
where $\langle 0|\frac{\alpha _{\mathrm{s}}}{\pi }G_{\mu \nu }^{A}\widetilde{%
G}^{A,\mu \nu }|\eta (p)\rangle $ is the matrix element appeared due to the $%
U(1)$ axial-anomaly. The\ quantities $h_{\eta }^{s(q)}$ can be expressed in
terms of the parameters $h_{s}$, $h_{q}$ and mixing angle $\varphi $%
\begin{equation}
h_{\eta }^{s}=-h_{s}\sin \varphi ,\ h_{\eta }^{q}=h_{q}\cos \varphi
\end{equation}%
which modifies Eq.\ (\ref{eq:ME3})
\begin{equation}
\langle 0|J^{\eta }|\eta (p)\rangle =-\frac{H^{\eta }}{2m_{s}},
\label{eq:ME3A}
\end{equation}%
where we introduce the short-hand notation $H^{\eta }=\left( h_{q}\cos
^{2}\varphi +h_{s}\sin ^{2}\varphi \right) $. In calculations we employ the
numerical values of $h_{q}$ and $h_{s}$ (in $\mathrm{GeV}^{3}$)
\begin{equation}
h_{q}=0.0016\pm 0.004,\ h_{s}=0.087\pm 0.006
\end{equation}%
extracted from analysis of experimental data. The same phenomenological
analyses predict $\ \varphi =39.3^{\circ }\pm 1.0^{\circ }$.

Then the physical side of the sum rule can be recast into the form%
\begin{equation}
\Pi ^{\mathrm{Phys}}(p)=-\frac{H^{\eta }f_{a_{0}}m_{a_{0}}}{2m_{s}}\frac{%
m^{2}}{(p^{2}-m^{2})^{2}}+...,  \label{eq:PhysSide3}
\end{equation}%
where $m^{2}=(m_{a_{0}}^{2}+m_{\eta }^{2})/2$.

In the last equality we take into account that $p=p^{\prime }$ and $q=0$ ,
which is required when considering a vertex composed of a tetraquark and two
conventional mesons \cite{Agaev:2016dev}. In the case of vertices containing
only ordinary mesons calculation of the corresponding strong coupling can be
performed in the context of the LCSR method's full version: the limit $q=0$
is known there as the soft approximation. For tetraquark-meson-meson
vertices the full LCSR method reduces to its soft approximation, which is
only way to compute the strong couplings. Therefore, we use here technical
tools elaborated in the soft approximation by bearing in mind that in our
case this is only available approach to evaluate $g_{a\eta \pi }.$ In the
limit $q=0$ the correlation function $\Pi ^{\mathrm{Phys}}(p)$ depends on a
variable $p^{2}$, as a result we have to fulfil the one-variable Borel
transformation which yields%
\begin{equation}
\mathcal{B}\Pi ^{\mathrm{Phys}}(p)=-\frac{H^{\eta }f_{a_{0}}m_{a_{0}}m^{2}}{%
2m_{s}}\frac{e^{-m^{2}/M^{2}}}{M^{2}}+\ldots .  \label{eq:BorelTr2}
\end{equation}%
We proceed by computing the QCD side of the sum rule. It is easily seen that
$J_{q}(x)$ does not contribute to the correlation function $\Pi (p,q)$.
Indeed, by substituting the current $J_{q}(x)$ into Eq.\ (\ref{eq:CF2}) and
performing contractions of the $\overline{u}u$ and $\overline{d}d$ fields
from $J_{q}(x)$ with relevant parts of $J(x)$ we get apart from light $u,\ d$%
-quark propagators matrix elements of the local operators $\overline{s}%
\Gamma _{i}s$ (here, $\Gamma ^{j}=\mathbf{1,\ }\gamma _{5},\ \gamma
_{\lambda },\ i\gamma _{5}\gamma _{\lambda },\ \sigma _{\lambda \rho }/\sqrt{%
2}$ is the full set of Dirac matrices) sandwiched between the $\pi ^{0}$ and
vacuum
\begin{equation*}
\langle \pi ^{0}|\overline{s}(0)\Gamma _{i}s(0)|0\rangle ,
\end{equation*}%
which are identically equal to zero. In other words, only $-\sin \varphi
J_{s}(x)$ component of the $\eta $ meson's current contributes to the
correlation function $\Pi (p,q)$.

After some manipulations we get%
\begin{eqnarray}
&&\Pi ^{\mathrm{OPE}}(p)=\sin \varphi \int d^{4}xe^{ip\cdot x}\frac{\epsilon
\widetilde{\epsilon }}{\sqrt{2}}\left\{ \left[ \gamma _{5}\widetilde{S}%
_{s}^{ib}(x)\gamma _{5}\right. \right.  \notag \\
&&\left. \left. \times \widetilde{S}_{s}^{ei}(-x)\gamma _{5}\right] _{\alpha
\beta }\left( \langle \pi ^{0}|\overline{u}_{\alpha }^{a}u_{\beta
}^{d}|0\rangle -\langle \pi ^{0}|\overline{d}_{\alpha }^{a}d_{\beta
}^{d}|0\rangle \right) \right\} ,  \label{eq:OPE2A}
\end{eqnarray}%
where $\alpha $ and $\beta $ are spinor indices.

Calculations of the correlation function in accordance with recipes
described in a rather detailed form in Ref.\ \cite{Agaev:2016dev} reveal
that the matrix elements of the pion which contributes to $\Pi ^{\mathrm{OPE}%
}(p)$ are $\langle 0|\overline{u}i\gamma _{5}u|\pi ^{0}\rangle $ and $%
\langle 0|\overline{d}i\gamma _{5}d|\pi ^{0}\rangle $ given, for example, in
the form
\begin{equation}
\sqrt{2}\langle 0|\overline{u}i\gamma _{5}u|\pi ^{0}\rangle =f_{\pi }\mu
_{\pi },~\ \ \mu _{\pi }=-\frac{2\langle \overline{q}q\rangle }{f_{\pi }^{2}}%
.  \label{eq:MatEl3}
\end{equation}%
In Eq.\ (\ref{eq:MatEl3}) $f_{\pi }$ and $\langle \overline{q}q\rangle $ are
the pion decay constant and the quark vacuum condensate, respectively. Then
the Borel transform of $\mathcal{B}\Pi ^{\mathrm{OPE}}(p)=\Pi ^{\mathrm{OPE}%
}(M^{2})$ which is necessary to derive the sum rule reads%
\begin{eqnarray}
&&\Pi ^{\mathrm{OPE}}(M^{2})=-\frac{f_{\pi }\mu _{\pi }}{16\pi ^{2}}\sin
\varphi \int_{4m_{s}^{2}}^{\infty }dsse^{-s/M^{2}}  \notag \\
&&-\sin \varphi \left[ \frac{f_{\pi }\mu _{\pi }}{16}\langle \frac{\alpha
_{s}G^{2}}{\pi }\rangle -\frac{f_{\pi }\mu _{\pi }m_{s}}{6}\langle \overline{%
s}s\rangle \right] .  \label{eq:OPE3}
\end{eqnarray}%
Equating the Borel transforms $\mathcal{B}\Pi ^{\mathrm{Phys}}(p)$ and $\Pi
^{\mathrm{OPE}}(M^{2})$ we get the unsubtracted sum rule. But the sum rule
applicable to evaluate $g_{a\eta \pi }$ can be obtained only after
subtracting the contributions of excited states and continuum. In the soft
approximation an additional problem in this procedure is connected with
contributions to $\mathcal{B}\Pi ^{\mathrm{Phys}}(p)$ of excited states,
some of which even after Borel transformation remain unsuppressed \ \cite%
{Belyaev:1994zk}, and should be removed by applying the operator $\mathcal{P}%
(M^{2},m^{2})$ (see, Ref. \cite{Ioffe:1983ju})
\begin{equation}
\mathcal{P}(M^{2},m^{2})=\left( 1-M^{2}\frac{d}{dM^{2}}\right)
M^{2}e^{m^{2}/M^{2}}.  \label{eq:Operator}
\end{equation}%
As a result we derive our final sum rule for the strong coupling
\begin{equation}
g_{a\eta \pi }=-\frac{2m_{s}}{H^{\eta }f_{a_{0}}m_{a_{0}}m^{2}}\mathcal{P}%
(M^{2},m^{2})\Pi ^{\mathrm{OPE}}(M^{2},s_{0}),  \label{eq:SumRule1}
\end{equation}%
where $\Pi ^{\mathrm{OPE}}(M^{2},s_{0})$ is given by Eq.\ (\ref{eq:OPE3})
where the upper limit of the integral $\infty $ is replaced by $s_{0}$.

The decay process $a_{0}(980)\rightarrow K^{+}K^{-}$ is investigated by the
same manner. The differences here are connected with the correlation
function
\begin{equation}
\Pi _{K}(p,q)=i\int d^{4}xe^{ip\cdot x}\langle K^{+}(q)|\mathcal{T}%
\{J^{K^{-}}(x)J^{^{\dagger }}(0)\}|0\rangle ,  \label{eq:CF4}
\end{equation}%
with the interpolating current $J^{K^{-}}(x)$
\begin{equation}
J^{K^{-}}(x)=\overline{u}^{i}(x)i\gamma _{5}s^{i}(x),  \label{eq:Current4}
\end{equation}%
and also the matrix element of the $K$ mesons%
\begin{equation}
\langle 0|\overline{u}i\gamma _{5}s|K^{-}(p)\rangle =\frac{f_{K}m_{K}^{2}}{%
m_{s}}.  \label{eq:MatEl4}
\end{equation}%
In Eq.\ (\ref{eq:MatEl4}) $m_{K}$ and $f_{K}$ are the $K^{\pm }$ mesons'
mass and decay constant, respectively. After relevant replacements the
phenomenological side of sum rule is obtained from Eq.\ (\ref{eq:PhysSide2}%
), whereas for $\Pi _{K}^{\mathrm{OPE}}(p,q)$ we get
\begin{eqnarray}
&&\Pi _{K}^{\mathrm{OPE}}(p,q)=i^{2}\int d^{4}xe^{ip\cdot x}\frac{\epsilon
\widetilde{\epsilon }}{\sqrt{2}}\left[ \gamma _{5}\widetilde{S}%
_{s}^{ib}(x)\gamma _{5}\widetilde{S}_{u}^{di}(-x)\gamma _{5}\right] _{\alpha
\beta }  \notag \\
&&\times \langle K^{+}(q)|\overline{u}_{\alpha }^{a}(0)s^{e}(0)|0\rangle .
\label{eq:OPE3A}
\end{eqnarray}%
The following operations are standard manipulations, therefore we write down
only the final sum rule for the strong coupling $g_{a_{0KK}}$
\begin{equation}
g_{aKK}=\frac{m_{s}}{m_{a_{0}}f_{a_{0}}m_{K}^{2}f_{K}\widetilde{m}^{2}}%
\mathcal{P}(M^{2},\widetilde{m}^{2})\Pi _{K}^{\mathrm{OPE}}(M^{2},s_{0}),
\label{eq:SumRule2}
\end{equation}%
where $\widetilde{m}^{2}=(m_{a_{0}}^{2}+m_{K}^{2})/2$ and
\begin{eqnarray}
&&\Pi _{K}^{\mathrm{OPE}}(M^{2},s_{0})=-\frac{f_{K}m_{K}^{2}}{16\sqrt{2}\pi
^{2}m_{s}}\int_{4m_{s}^{2}}^{s_{0}}dsse^{-s/M^{2}}  \notag \\
&&+\frac{f_{K}m_{K}^{2}}{16\sqrt{2}m_{s}}\langle \frac{\alpha _{s}G^{2}}{\pi
}\rangle -\frac{f_{K}m_{K}^{2}}{12\sqrt{2}}\left( 2\langle \overline{u}%
u\rangle -\langle \overline{s}s\rangle \right).  \label{eq:OPE4}
\end{eqnarray}

For the strong couplings $g_{K^{\ast }K\pi }$ and $g_{K^{\ast }K^{0}\pi
^{0}} $ we obtain:%
\begin{equation}
g_{K^{\ast }K\pi }=\frac{m_{s}}{m_{K^{\ast }}f_{K^{\ast
}}m_{K}^{2}f_{K}m_{1}^{2}}\mathcal{P}(M^{2},m_{1}^{2})\Pi _{1}^{\mathrm{OPE}%
}(M^{2},s_{0}),  \label{eq:SumRule3}
\end{equation}%
and
\begin{equation}
g_{K^{\ast }K^{0}\pi ^{0}}=\frac{m_{s}}{m_{K^{\ast }}f_{K^{\ast
}}m_{K^{0}}^{2}f_{K^{0}}m_{2}^{2}}\mathcal{P}(M^{2},m_{2}^{2})\Pi _{2}^{%
\mathrm{OPE}}(M^{2},s_{0}),  \label{eq:SumRule4}
\end{equation}%
where $m_{1}^{2}=(m_{K^{\ast }}^{2}+m_{K}^{2})/2$ and $m_{2}^{2}=(m_{K^{\ast
}}^{2}+m_{K^{0}}^{2})/2$, respectively. The correlation functions in Eqs.\ (%
\ref{eq:SumRule3}) and (\ref{eq:SumRule4}) are given by the expressions%
\begin{eqnarray}
&&\Pi _{1}^{\mathrm{OPE}}(M^{2},s_{0})=-\frac{f_{\pi }\mu _{\pi }}{16\pi ^{2}%
}\int_{m_{s}^{2}}^{s_{0}}dsse^{-s/M^{2}}  \notag \\
&&-\frac{f_{\pi }\mu _{\pi }}{16}\langle \frac{\alpha _{s}G^{2}}{\pi }%
\rangle +\frac{f_{\pi }\mu _{\pi }m_{s}}{12}\left( 2\langle \overline{u}%
u\rangle -\langle \overline{s}s\rangle \right) ,  \label{eq:OPE5}
\end{eqnarray}%
and $\Pi _{2}^{\mathrm{OPE}}(M^{2},s_{0})=\Pi _{1}^{\mathrm{OPE}%
}(M^{2},s_{0})/\sqrt{2}.$

Sum rules obtained for the strong couplings \ $g_{a\eta \pi }$, $g_{aKK}$, $%
g_{K^{\ast }K\pi }$ and $g_{K^{\ast }K^{0}\pi ^{0}}$ will be used to
determine the partial decay widths of the mesons $a_{0}(980)$ and $%
K_{0}^{\ast }(800)$.


\section{Numerical analysis}

\label{sec:NRes}

In numerical computations of the strong couplings for the quark and gluon
condensates we utilize their values presented in Eq.\ (\ref{eq:Param}).
Apart from these parameters we also employ the masses and decay constants of
the $\pi $ and $K$ mesons: for the pion $m_{\pi ^{\pm }}=139.57061\pm
0.00024\ \mathrm{MeV}$, $m_{\pi ^{0}}=134.9770\pm 0.0005\ \mathrm{MeV}$ and $%
f_{\pi }=131\ \mathrm{MeV}$ and for the $K$ meson $m_{K^{\pm }}=493.677\pm
0.016\ \mathrm{MeV}$, $m_{K^{0}}=497.611\pm 0.013\ \mathrm{MeV}$ and $%
f_{K}=155.72\ \mathrm{MeV.}$

We have employed the different working regions for the Borel parameter $%
M^{2} $ and continuum threshold $s_{0}$ when considering decays of the $%
a_{0}(980)$ and $K_{0}^{\ast }(800)$ mesons: these windows have been chosen
in accordance with standard constraints of the sum rule computations
explained in the section \ref{sec:Masses}. For the strong couplings $%
g_{a\eta \pi }$ and $g_{aKK}$ the Borel and continuum threshold parameters
are varied within the limits
\begin{equation}
M^{2}\in \lbrack 1.1-1.4]\ \mathrm{GeV}^{2},\ s_{0}\in \lbrack 1.9-2.1]\
\mathrm{GeV}^{2}.  \label{eq:Wind3}
\end{equation}%
The corresponding sum rules lead to the following predictions (in units of $%
\mathrm{GeV}^{-1}$)%
\begin{equation}
g_{a\eta \pi }=5.36\pm 1.41\ ,\ \ g_{aKK}=9.10\pm 2.76.  \label{eq:SCoupl1}
\end{equation}

\begin{widetext}

\begin{figure}[h!]
\begin{center}
\includegraphics[totalheight=6cm,width=8cm]{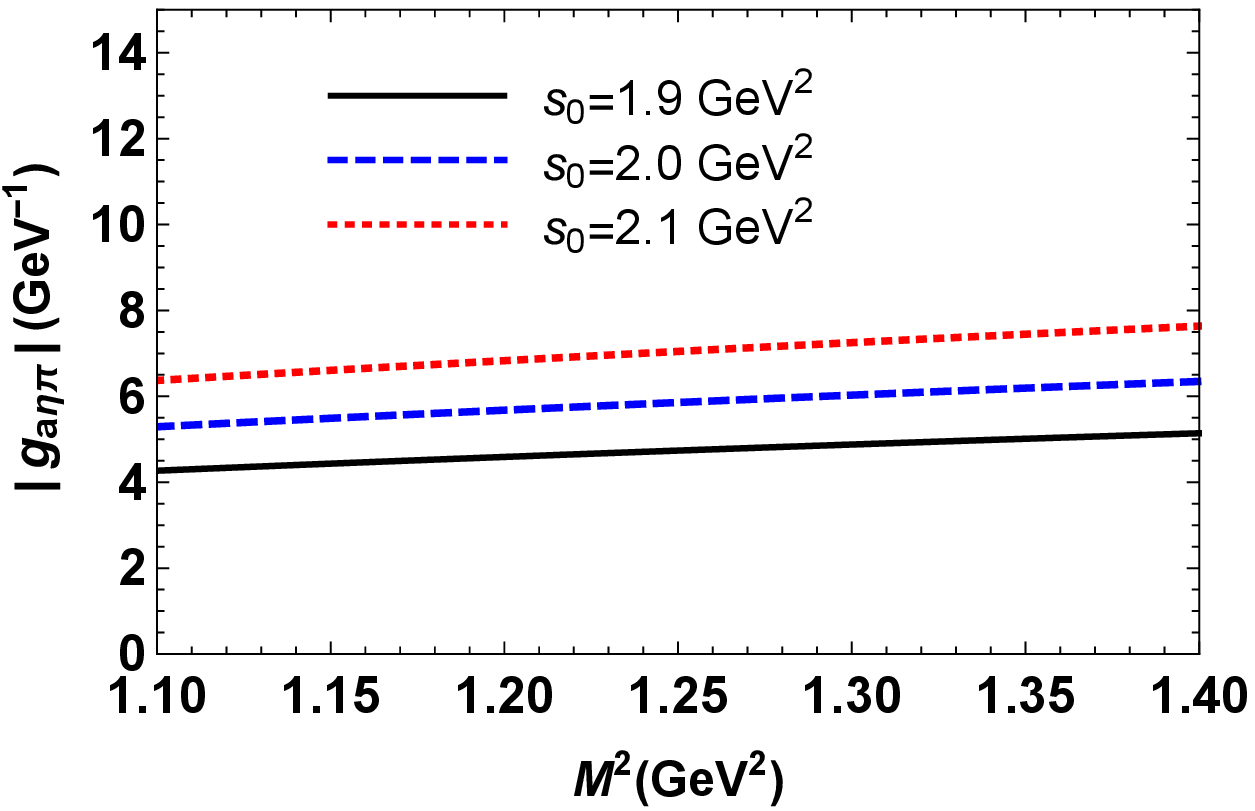}\,\,
\includegraphics[totalheight=6cm,width=8cm]{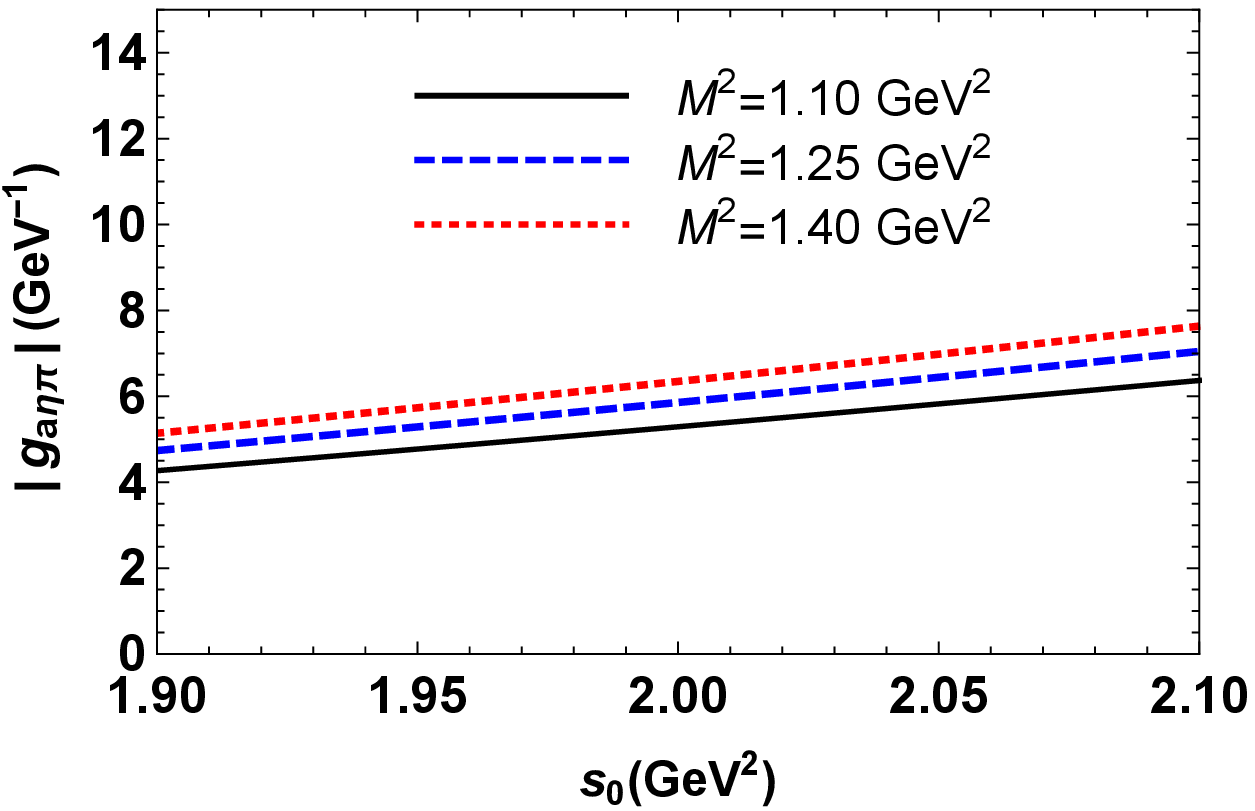}
\end{center}
\caption{ The strong coupling $g_{a_0 \eta \pi}$ as a function of the Borel parameter $M^2$
(left panel), and of the continuum threshold $s_0$ (right panel)}.
\label{fig:StCoupl}
\end{figure}

\end{widetext}

It is known that a stability of the obtained results on $M^{2}$ and $s_{0}$
is one of the important constraints imposed on sum rule computations. As an
example, in Fig.\ \ref{fig:StCoupl} we plot the coupling $g_{a\eta \pi }$ as
a function of $M^{2}$ and $s_{0}$. It is evident that $g_{a\eta \pi }$
depends on $M^{2}$ and $s_{0}$ , which generates essential part of
uncertainties in the evaluated quantities. It is also seen that these
ambiguities do not exceed $\sim 30\%$ of the central values which is acceptable
for the sum rules computations.

For the partial decay width of the processes $a_{0}(980)\rightarrow \eta \pi
^{0}$ and $a_{0}(980)\rightarrow K^{+}K^{-}$ we get%
\begin{eqnarray}
\Gamma \left[ a_{0}(980)\rightarrow \eta \pi ^{0}\right] &=&50.57\pm 13.87\
\mathrm{MeV,}  \notag \\
\Gamma \left[ a_{0}(980)\rightarrow K^{+}K^{-}\right] &=&11.44\pm 3.76\
\mathrm{MeV.}  \label{eq:DW1}
\end{eqnarray}%
The total width of the meson $a_{0}(980)$ is formed mainly due to the decay
channels \ $a_{0}(980)\rightarrow \eta \pi ^{0}$ and $a_{0}(980)\rightarrow
K^{+}K^{-}$: we assume that $P-$wave decays do not modify it considerably.
Therefore it seems reasonable to compare $\Gamma _{\mathrm{th}.}=62.01\pm
14.37\ \ \mathrm{MeV}$ which is the sum of two partial decay widths with the
available information on $\Gamma _{\exp .}=50-100\ \ \mathrm{MeV}$ noting a
full overlap of these results. As we have noted above, experimental data for
the total width of the light scalar mesons suffer from large uncertainties.
Therefore, we can state that our theoretical prediction does not contradict
to the present-day experimental data.

The strong decays of the meson $K_{0}^{\ast }(800)$ can be analyzed in the
same way. In the case of the $K_{0}^{\ast }(800)$ meson's decays we use
\begin{equation}
M^{2}\in \lbrack 0.8-1.0]\ \mathrm{GeV}^{2},\ \ s_{0}\in \lbrack 1.2-1.5]\
\mathrm{GeV}^{2},  \label{eq:RegionsK}
\end{equation}%
and find for the strong couplings (in $\mathrm{GeV}^{-1}$)
\begin{equation}
g_{K^{\ast }K\pi }=19.46\pm 5.64\ ,\ \ \ g_{K^{\ast }K^{0}\pi ^{0}}=13.47\pm
3.91.  \label{eq:StCoupl2}
\end{equation}%
The partial decay widths are equal to
\begin{eqnarray}
\Gamma \left[ K_{0}^{\ast }(800)\rightarrow K^{+}\pi ^{-}\right]
&=&270.39\pm 78.42\ \mathrm{MeV,}  \notag \\
\Gamma \left[ K_{0}^{\ast }(800)\rightarrow K^{0}\pi ^{0}\right]
&=&130.69\pm 37.91\ \mathrm{MeV.}  \label{eq:DW2}
\end{eqnarray}%
Then using these two decay modes for the total width of $K_{0}^{\ast }(800)$
we get $\Gamma _{\mathrm{th}.}=401.1\pm 87.1\ \ \mathrm{MeV}$. Experimental
data borrowed from Refs. \ \cite{Ablikim:2010ab,Ablikim:2010kd} predicts $%
\Gamma _{\exp .}=449\pm 156_{\ -81}^{\ +144}\ \ \mathrm{MeV}$ and $\Gamma
_{\exp .}=512\pm 80_{\ -44}^{\ +92}\ \ \mathrm{MeV}$, respectively, which
have rather imprecise nature. It is seen that our result is compatible with these data.  

\section{Discussion and concluding notes}

\label{sec:CRel}

Investigation of the scalar mesons $a_{0}(980)$ and $K_{0}^{\ast }(800)$ by
modeling them as diquark-antidiquarks carried out in the present work has
allowed us to explore the suggestion about exotic nature of these
resonances. Using the well-known QCD sum rule method we have calculated
their masses and total widths. To this end, we have employed the
interpolating currents $J(x)$ and $J^{K^{\ast }}(x)$ defined by Eqs.\ (\ref%
{eq:Current1}) and (\ref{eq:Current2}), respectively.

Our investigation has demonstrated that single currents can be successfully
applied to interpolate the light scalar mesons. In this point we do not
agree with Ref.\ \cite{Chen:2007xr}, in which the authors excluded single
interpolating currents as ones that do not lead to reliable predictions. An
accuracy of theoretical calculations performed in our work exceeds an
accuracy of similar computations in Ref.\ \cite{Chen:2007xr}. Thus, in our
study we have taken into account not only terms up to dimension ten instead
of eight, but also used in calculations more precise expression for the
quark propagator. It is possible that conclusion made Ref.\ \cite%
{Chen:2007xr} is connected with these circumstances.

Our result for the mass of the $a_{0}(980)$ agrees with experimental data.
Its total width evaluated using two $S-$wave dominant strong decay channels
is also in accord with the data, because our result lies entirely in the
experimental region $\Gamma _{\exp .}=50-100$ $\mathrm{MeV.}$ The situation
with experimental information on the parameters of the $K_{0}^{\ast }(800)$
meson is worse than in the case of \ $a_{0}(980)$. Thus, available data on
both the mass and total width of this scalar meson in rather imprecise and
suffers from large uncertainties. The predictions obtained in the present
work do not contradict to last experimental measurements, nevertheless
reliable conclusions can be made only on basis of a more precise
experimental information.

\section{Acknowledgments}

K.A. and H.S. thank TUBITAK for the partial financial support provided under
Grant No. 115F183.

\end{document}